\documentstyle[psfig,12pt]{article}
\def\TL{\hfil$\displaystyle{##}$}
\def\TR{$\displaystyle{{}##}$\hfil}
\def\TC{\hfil$\displaystyle{##}$\hfil}
\def\TT{\hbox{##}}
\def\seqalign#1#2{\vcenter{\openup1\jot
  \halign{\strut #1\cr #2 \cr}}}

\def\comment#1{}
\def\fixit#1{}

\def\tf#1#2{{\textstyle{#1 \over #2}}}

\def\mop#1{\mathop{\rm #1}\nolimits}

\def\tr{\mop{tr}}
\def\Disc{\mop{Disc}}

\def\overleftrightarrow#1{\vbox{\ialign{##\crcr
     $\leftrightarrow$\crcr\noalign{\kern-0pt\nointerlineskip}
     $\hfil\displaystyle{#1}\hfil$\crcr}}}

\def\lsim{\mathrel{\mathstrut\smash{\ooalign{\raise2.5pt\hbox{$<$}\cr\lower2.5pt\hbox{$\sim$}}}}}
\def\gsim{\mathrel{\mathstrut\smash{\ooalign{\raise2.5pt\hbox{$>$}\cr\lower2.5pt\hbox{$\sim$}}}}}

\def\slashed#1{\ooalign{\hfil\hfil/\hfil\cr $#1$}}

\def\sqr#1#2{{\vcenter{\vbox{\hrule height.#2pt
         \hbox{\vrule width.#2pt height#1pt \kern#1pt
            \vrule width.#2pt}
         \hrule height.#2pt}}}}
\def\square{\mathop{\mathchoice\sqr56\sqr56\sqr{3.75}4\sqr34\,}\nolimits}

\def\href#1#2{#2}  

\def\lbldef#1#2{\expandafter\gdef\csname #1\endcsname {#2}}
\def\eqn#1#2{\lbldef{#1}{(\ref{#1})}%
\begin{equation} #2 \label{#1} \end{equation}}
\def\eqalign#1{\vcenter{\openup1\jot
    \halign{\strut\span\TL & \span\TR\cr #1 \cr
   }}}

\def\comment#1{  \begin{raggedright}{\tt [#1]}\end{raggedright}}
\begin{document}
\baselineskip=15.5pt
\pagestyle{plain}
\setcounter{page}{1}
\renewcommand{\thefootnote}{\fnsymbol{footnote}}

\begin{titlepage}

\begin{flushright}
HUTP-99/A053 \\
hep-th/9910117
\end{flushright}
\vfil

\begin{center}
{\huge Non-conformal examples} \\[8pt]
{\huge of AdS/CFT}
\end{center}

\vfil
\begin{center}
{\large Steven S. Gubser}
\end{center}

$$\seqalign{\span\TL & \span\TT}{
& Lyman Laboratory of Physics, Harvard University, Cambridge, MA
02138, USA 
}$$
\vfil

\begin{center}
{\large Abstract}
\end{center}

\noindent
 Asymptotically anti-de Sitter spacetimes with Poincar\'e invariance along
the boundary can describe, via the AdS/CFT correspondence, either relevant
deformations of a conformal field theory or non-conformal vacuum states.  I
consider examples of both types constructed in the framework of
five-dimensional gauged supergravity.  I explain the proof and motivation
of a gravitational ``c-theorem'' which is independent of dimension.  I show
how one class of examples can be elevated to ten-dimensional geometries
involving distributions of parallel D3-branes.  For these cases some
peculiar properties of two-point functions emerge, and I close with
speculations on their physical origin.

\vfil
\begin{flushleft}
October 1999
\end{flushleft}
\end{titlepage}
\newpage
\section{Introduction}
\label{Introduction}

The AdS/CFT correspondence \cite{juanAdS,gkPol,witHolOne} (for a review see
\cite{MAGOO}) offers us the opportunity to study not only exactly conformal
field theories and their anti-de Sitter duals, but also deformations of
conformal field theories corresponding to distortions of the anti-de Sitter
geometry.  A subtlety that was realized early on is that these distortions
can represent either changes in the hamiltonian of the dual field theory or
changes in the physical state.  Which obtains in any particular case is
determined by the asymptotics of the various fields involved near the
boundary of anti-de Sitter space.  Resolving this ambiguity has been one of
the main obstacles to formulating the renormalization group (RG) equations
clearly in an AdS/CFT setting.  Very roughly, supergravity equations of
motion are second order, admitting two solutions---one solution being a
deformation of the hamiltonian and the other a change in the physical
state---but RG equations are first order, and any translation of RG into
AdS/CFT language must somehow specify which supergravity solution the RG
picks out.  Conventionally we think of RG as acting on the hamiltonian
rather than the state, so we should be looking for a way to track
deformations of the hamiltonian with the state held fixed.

An improved understanding of RG is one motivation of the work I will
describe.  The result which touches most directly on RG is a gravitational
``c-theorem,'' which I will exhibit in section~\ref{cTheorem} after some
preliminaries to set conventions and make slightly more precise the
distinction between states and deformations.  I will then move on in
section~\ref{ItsADuck} to an interesting class of examples which represent
states in the Coulomb branch of ${\cal N}=4$ super-Yang-Mills theory.  The
supergravity geometries can be described wholly in terms of
five-dimensional ${\cal N}=8$ gauged supergravity, but the physics is
clearest from the distribution of D3-branes in ten dimensions from which
these geometries descend.  They provide a simple example of consistent
truncation.  They also exhibit two surprising features: 1) the distribution
of D3-branes is not always positive definite, and 2) the spectrum of
supergravity excitations is often gapped or even discrete.  The spectrum
can be obtained through an analysis of two-point functions, and this will
be the topic of section~\ref{TwoPoint}.  This talk is mostly based on
collaborative work with D.~Freedman, K.~Pilch, and N.~Warner
\cite{fgpwOne,fgpwTwo}.

\section{A c-theorem from gravity}
\label{cTheorem}

In order to have a clear methodology for computing physical
quantities such as correlators, I will restrict attention to spaces which
are asymptotically anti-de Sitter.  Furthermore, all my examples will have
3+1-dimensional Poincar\'e invariance, so that the five-dimensional
geometry can be written in the form
  \eqn{FiveGeometry}{
   ds^2 = e^{2A(r)} (dt^2 - dx_1^2 - dx_2^2 - dx_3^2) - dr^2 \ .
  }
 With this choice of coordinates, $A = r/L$ corresponds to pure $AdS_5$ (or
more precisely a Poincar\'e patch of $AdS_5$), with $R_{\mu\nu} = {4 \over
L^2} g_{\mu\nu}$.  If $A$ is not linear, then there must be some matter
fields providing an $r$-dependent stress-energy tensor which supports the
geometry.  In order to have Poincar\'e invariance, essentially the only
possibility is that the matter fields should be scalars.  For the space to
be asymptotically $AdS_5$, we need the scalars to approach constant values
near the boundary.  Those constant values must represent an extremum of the
scalar potential, and by convention we will say the scalars vanish there.
Near the boundary one can use a linearized approximation to the scalar
equations of motion: for one scalar, $(\square+m^2)\phi=0$.  The two
independent solutions map to deformations and vacuum states in the
following way:
  \eqn{StateVacuum}{\seqalign{\span\TL & \span\TR & \quad\ \span\TC\quad\ & 
    \span\TR}{
   \phi &\sim e^{-\Delta_- r/L} & \longleftrightarrow & 
    {\cal L} \to {\cal L} + e^{\Delta_- r/L} \phi {\cal O}  \cr
   \phi &\sim e^{-\Delta_+ r/L} & \longleftrightarrow &
    \langle {\cal O} \rangle = e^{\Delta_+ r/L} \phi
  }}
 where $\Delta_\pm = 2 \pm \sqrt{4+(mL)^2}$ and ${\cal O}$ is the color
singlet operator (of dimension $\Delta_+$) in the conformal field theory
which is dual to $\phi$.  In words, the more singular asymptotics of $\phi$
corresponds to adding a source for ${\cal O}$ to the lagrangian, while the
less singular solution corresponds to giving ${\cal O}$ a VEV.  These are
dual concepts in the sense of Legendre transforms.  It is possible in
certain circumstances for the roles of $\Delta_+$ and $\Delta_-$ to be
interchanged \cite{klebWitTwo}.  This talk will focus on deformations and
states of ${\cal N}=4$ super-Yang-Mills, where the conventional
identification of the more singular solution as a deformation and the less
singular one as a VEV is correct for all the fields.
   
There is a general intuition in quantum field theory that there is a
thinning out of degrees of freedom as one passes from the ultraviolet to
the infrared.  This holds equally whether the energy scales of the theory
arise from the physical state (as in spontaneous symmetry breaking) or from
the dynamics of the lagrangian (as in confinement).  A first step, then, in
finding the meaning of the renormalization group in the language of AdS/CFT
is to see how and whether one can quantify the thinning out process.

In two dimensions there is a celebrated theorem of Zamolodchikov
\cite{Zamolodchikov:1986gt} to the effect that renormalization group flows
follow steepest descent trajectories of a so-called ``c-function'' of the
couplings in the hamiltonian.  At fixed points of the renormalization group
the c-function coincides with the central charge in the Virasoro algebra,
or equivalently with the coefficient $c$ in the anomalous one-point
function
  \eqn{TTwoDim}{
   \langle T^\mu_\mu \rangle_{g_{\mu\nu}} = -{c \over 24 \pi} R
  }
 where $g_{\mu\nu}$ is an external metric and $R$ is its scalar curvature.
The analogous expression in four dimensions, in the notation of
\cite{Anselmi:1997am}, is 
  \eqn{TFourDim}{
   \langle T^\mu_\mu \rangle_{g_{\mu\nu}} =
    {c \over 16 \pi^2} W_{\mu\nu\rho\sigma}^2 -
    {a \over 16 \pi^2} \tilde{R}_{\mu\nu\rho\sigma}
     \tilde{R}^{\mu\nu\rho\sigma} \ .
  }
 The first term is the square of the Weyl tensor and the second is the
topological Euler density.  It has been conjectured by Cardy
\cite{Cardy:1988cw} that any renormalization group flow which begins and
ends at a conformal fixed point must have a lower value of $a$ in the
infrared than the ultraviolet.

On the other hand, it has been shown \cite{hs} using the prescription of
\cite{gkPol,witHolOne} that a pure $AdS_5$ geometry (that is,
\FiveGeometry\ with $A(r) = r/L$) leads to a one-point function of the form
\TFourDim\ with
  \eqn{hsAandC}{
   a = c = {\pi^2 \over \kappa_5^2 A'(r)^3} \ .
  }
 In arbitrary even boundary dimension $d$ a similar expression emerges from
the analysis \cite{hs} with all anomaly coefficients in a fixed ratio and
proportional to $1/A'(r)^{d-1}$, which again is constant since $A(r)$ is
linear.

The remarks of the last two paragraphs suggest a natural ``c-theorem'' in
AdS/CFT: can it be shown that in a geometry where $A(r)$ approaches linear
behavior both as $r \to +\infty$ (near the boundary of $AdS_{d+1}$) and as
$r \to -\infty$ (in the deep interior), the limiting value of
$1/A'(r)^{d-1}$ is smaller in the deep interior than near the boundary?
The answer is yes for a very straightforward reason: inspecting the
curvature tensors for the metric \FiveGeometry, one finds that
  \eqn{cProof}{
   -(d-1) A'' = R^t_t - R^r_r = G^t_t - G^r_r = 
    \kappa_d^2 (T^t_t - T^r_r) \geq 0
  }
 where in the second to last step we have used Einstein's equations, and in
the last step we have assumed the null energy condition: $T_{\alpha\beta}
\xi^\alpha \xi^\beta \geq 0$ for any null vector $\xi^\alpha$.  This energy
condition is obeyed for all the matter fields in gauged supergravity, and
to my knowledge also in string theory (I exclude from consideration
orientifold planes, which are non-dynamical in the sense that their
location is fixed).  It should be noted that all curvature tensors as well
as the stress tensors in \cProof\ are $d$+1-dimensional quantities relating
to the supergravity dual picture, whereas in \TFourDim\ we were discussing
$d$-dimensional quantities relating to the boundary quantum field theory.
An equivalent form of \hsAandC\ was considered independently in
\cite{GPPZ}, and its monotonicity was checked for the non-supersymmetric
flows studied there.

A compelling point of evidence in favor of the definition \hsAandC\ is that
it leads to the correct central charges for supersymmetric RG flows between
conformal fixed points.  The field theory techniques for analyzing these
flows were developed in \cite{LeighStrassler}.  AdS/CFT examples have been
studied in \cite{klebWitOne,gEin,KPW,klm,fgpwOne}.  I will focus on the
case worked out in complete detail in \cite{fgpwOne}, namely ${\cal N}=4$
super-Yang-Mills theory deformed to an ${\cal N}=1$ theory by adding a mass
for one of the adjoint chiral superfields.  This has already been discussed
in the talk by N.~Warner, so I will be brief.

The superpotential for the field theory in question is
  \eqn{superW}{
   W = \tr \phi_3 [\phi_1,\phi_2] + {m \over 2} \tr \phi_3^2 \ .
  }
 Using the techniques of \cite{LeighStrassler}, the infrared limit can be
shown to be a strongly interacting fixed point with a quartic
superpotential proportional to $\tr {}[\phi_1,\phi_2]^2$.  The anomaly
coefficients $c$ and $a$ can be computed from anomalies in the divergence
of the $R$-current: this is on account of the fact that in ${\cal N}=1$
supersymmetry, $\partial_\mu R^\mu$ and $T^\mu_\mu$ are in the same
multiplet.  The result is that $c=a$ both in the infrared and the
ultraviolet, and $c_{UV} = N^2/4$ while $c_{IR} = {27 \over 32} c_{UV}$.
't~Hooft anomaly matching is required to obtain the latter result.  These
predictions are precisely matched by \hsAandC\ applied to the dual
supergravity geometry \cite{fgpwOne}, which was described in detail by
N.~Warner earlier in this conference.  The first check that AdS/CFT
predicted the correct value for $c_{IR}/c_{UV}$ was carried out in
\cite{gEin}.  Subsequent work on the subject includes \cite{klm,fgpwOne}.

In the case of supersymmetric flows, it was possible to express the
c-function as an explicit function of the supergravity scalars
\cite{fgpwOne}: all $r$-dependence in $C(r)$ arose strictly through the
scalar profiles.  The scalars' evolution followed the gradient flow of an
appropriate power of the c-function.  This has very much the flavor of
Zamolodchikov's c-theorem, but it seemed to rely on supersymmetry.  However
it was recently shown \cite{dfgk} that any solution to the equations of
motion of 
  \eqn{NLSAction}{
   S=\int d^4x dr \, \sqrt{|\mbox{det}g_{\mu \nu}|}
    \left [ -\frac{1}{4} R + \frac{1}{2} G_{IJ} \partial^{\mu} \phi^J
    \partial_{\mu} \phi^I - V(\phi)
    \right ] \,,
  }
 of the form \FiveGeometry\ with scalar profiles depending only on $r$
could be obtained as solutions to the first order
equations\footnote{R.~Myers has also noted that this solution generating
technique \cite{MyersUnpublished} can be applied even in the absence of
supersymmetry.}
  \eqn{WeomsTwo}{ 
   {d\phi^I \over dr} = 
     {1 \over 2} \, G^{IJ} \, {\partial W(\phi) \over \partial\phi^J} \,,
    \qquad {dA \over dr}  = -{1 \over 3} W(\phi) \,, 
  } 
 where $W(\phi)$ is a ``superpotential'' (not to be confused with the
boundary field theory superpotential \superW) satisfying
  \eqn{VWFormTwo}{ 
   V(\phi) = {1 \over 8} \, G^{IJ} \, {\partial W(\phi) \over \partial \phi^I} \,
    {\partial W(\phi) \over \partial \phi^J} - {1 \over 3} W(\phi)^2\ .
  } 
 The point is that, given $V(\phi)$ \VWFormTwo\ can in principle be solved
for $W(\phi)$.  Thus the relation $C \propto 1/W(\phi)^3$, first derived in
\cite{fgpwTwo} for supersymmetric flows, is seen to extend to the
non-supersymmetric context \cite{dfgk}---only now $W(\phi)$ is not unique
but rather depends on as many constants of integration as there are
scalars.  This gives me some new hope that the RG equations might yet be
related explicitly to supergravity equations, with the aforementioned
constants of integration specifying what physical quantities are held fixed
in the renormalization.  However I should note the drawback that $W(\phi)$
is not a single-valued function of the $\phi^I$.  The graph of $W$ versus
$\phi^I$ is a piecewise smooth, codimension one hypersurface in the region
of $(W,\phi^I)$ space where $V(\phi) + {1 \over 3} W^2 > 0$.  It has
discontinuous slope at the boundary of this region: it ``reflects'' off the
boundary to form multiple sheets for the function $W(\phi)$.

\section{States on the Coulomb branch of ${\cal N}=4$ super-Yang-Mills}
\label{ItsADuck}

\begin{figure}
   \centerline{\psfig{figure=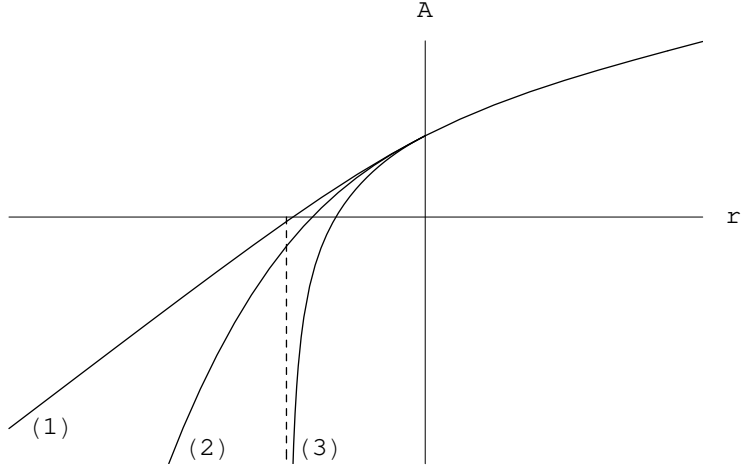,width=4in}}
    \caption{Possible behaviors for $A(r)$.  In case (1), constant negative
curvature is recovered asymptotically as $r\to-\infty$.  In case (2),
$A'(r)$ is unbounded below but $A$ is defined for all $r$.  In case (3),
$A(r) \to -\infty$ at finite $r$.}\label{figA}
  \end{figure}

The main geometrical fact we have learned from the previous section is that
the function $A(r)$ appearing in the metric \FiveGeometry\ is concave down
as a function of $r$: that is, $A''(r) \leq 0$.  The inequality is
saturated precisely for (locally) anti-de Sitter space.  There are then
three different possibilities for the behavior of $A(r)$ in geometries dual
to field theories with conformal invariance broken either by VEV's or
relevant deformations.  They are sketched in figure~\ref{figA}.

All three possibilities are realized in various contexts in the literature.
Case (1) occurs for the renormalization group flow discussed in the
previous section.  Case (3) is perhaps the most generic, and occurs in
various attempts to find a geometry dual to confining gauge theories
\cite{ksDil,gDil,GPPZGeneric}.  It has the pathology of a naked timelike
singularity at the radius where $A(r) \to -\infty$.  The borderline case
(2) has unbounded curvatures as $r \to -\infty$, but there are no genuine
curvature singularities inside the Poincar\'e patch.\footnote{It is a
largely open question what is the maximal analytic extension of the bulk
geometries discussed in this lecture.}  An example of (2) was obtained in
\cite{fgpwTwo} as a state on the Coulomb branch of ${\cal N}=4$
super-Yang-Mills, and we will discuss it below.  Earlier work on the
subject includes \cite{klt,klebWitTwo}, and the independent work of
\cite{ABKS} has some overlap with the results of \cite{fgpwTwo}.

The Coulomb branch of $SU(N)$ ${\cal N}=4$ super-Yang-Mills theory is
parametrized by VEV's of the scalar fields $X_I$, or equivalently the
relative positions of the $N$ D3-branes in the transverse six dimensions.
We should be able to describe these states in terms of an asymptotically
$AdS_5$ geometry with some scalar profiles specified near the boundary.
Instead of specifying directly the eigenvalues of the $X_I$, this approach
amounts to specifying the quantities $\tr X_{(I_1} X_{I_2} \ldots
X_{I_\ell)}$.  Actually these gauge-invariant traces do not wholly fix the
eigenvalues of the $X_I$, as we will see in examples.

The simplest case from the point of view of five-dimensional supergravity
is to give VEV's only to those scalars which are in the same $d=5$, ${\cal
N}=8$ supersymmetry multiplet as the graviton.  There are $42$ such
scalars, of which $20$ with $(mL)^2 = -4$ are dual to the operators $\tr
X_{(I} X_{J)}$.  They parametrize the coset $SL(6,{\bf R})/SO(6)$.  A
explicit discussion of the five-dimensional geometries in question was
given in the talk by N.~Warner: they all preserve 16 supersymmetries and a
$SO(n) \times SO(6-n)$ subgroup of $SO(6)$ (by convention $SO(1)$ is the
trivial group).  Here I will focus on the ten-dimensional origin of these
geometries and on the spectrum of supergravity excitations which they give
rise to.

One obvious point is that the ten-dimensional geometries must be
near-horizon limits of multi-center D3-brane solutions: that is,
  \eqn{MultiCenter}{\eqalign{
   ds_{10}^2 &= {1 \over \sqrt{H}} (dt^2 - dx_1^2 - dx_2^2 - dx_3^2) - 
    \sqrt{H} (dy_1^2 + dy_2^2 + \ldots + dy_6^2)  \cr
   H &= {L^4 \over N} \sum_{i=1}^N {1 \over |\vec{y}-\vec{y}_i|^4} \approx
        {L^4 \over N} \int d^6 y' 
         {\sigma(\vec{y}') \over |\vec{y}-\vec{y}'|^4} \ .
  }}
 Here $L^4 = {\kappa_{10} N \over 2\pi^{5/2}}$ and $\sigma(\vec{y})$ is a
continuous distribution normalized so that its integral is one.  (In the
supergravity approximation, discreteness of D3-branes is lost).  The $SO(n)
\times SO(6-n)$ symmetry of the five-dimensional supergravity solution must
translate into a symmetry of the distribution $\sigma(\vec{y})$.  It
emerged from the analysis of \cite{fgpwTwo} that the distributions in
question were
  \begin{eqnarray}
n=0:
&&\sigma= \delta^6(\vec y)\,,\nonumber\\
n=1:
&&\sigma =\tf{2}{\pi\ell} (1-\tf{y_1^2}{\ell^2})^{1/2}\,
\Theta(1-\tf{y_1^2}{\ell^2})\, \delta^{(5)}(\vec y_{2,3,4,5,6})
\,,\nonumber\\
n=2:
&&\sigma =\tf{1}{\pi\ell^2}\, 
\Theta(1-\tf{y_1^2}{\ell^2} -\tf{y_2^2}{\ell^2})\,
\delta^{(4)}(\vec y_{3,4,5,6})\,,\nonumber\\
n=3:
&&\sigma =\tf{1}{\pi^2\ell^3}\, 
(1-\sum_{i=1}^3\tf{y_i^2}{\ell^2})^{-1/2}\, 
\Theta(1-\sum_{i=1}^3\tf{y_i^2}{\ell^2})\,
\delta^{(3)}(\vec y_{4,5,6})\,,\label{d3distfun}\\
n=4:
&&\sigma =\tf{1}{\pi^2\ell^4}\, 
\delta(1-\sum_{i=1}^4\tf{y_i^2}{\ell^2})\,
\delta^{(2)}(\vec y_{5,6})\,,\nonumber\\
n=5:
&&\sigma =\tf{1}{2\pi^3\ell^5}\Big[-
(1-\sum_{i=1}^5\tf{y_i^2}{\ell^2})^{-3/2}\,
\Theta(1-\sum_{i=1}^5\tf{y_i^2}{\ell^2})\nonumber\\
&&\qquad\qquad\qquad  +2(1-\sum_{i=1}^5\tf{y_i^2}{\ell^2})^{-1/2}
\delta(1-\sum_{i=1}^5\tf{y_i^2}{\ell^2})\Big]\,
\delta(y_6)\,,\nonumber
  \end{eqnarray}
 where $\Theta(x)$ is $1$ for $x>0$ and $0$ for $x<0$.  It was observed in
\cite{cglp} that all of these distributions are limits (in the weak sense
of being convolved against any smooth function) of
  \begin{equation}
    n=6: \ \ \ 
     \sigma =\tf{1}{\pi^3\ell_1\cdots\ell_6}
      \delta'(1-\sum_{i=1}^6\tf{y_i^2}{\ell_i^2}) \ , \label{D3Gen}
  \end{equation}
 where $\delta'(x) = {d \over dx} \delta(x)$.  An obvious peculiarity of
the $n=5$ and $n=6$ distributions is that $\sigma(\vec{y})$ is not a
positive definite distribution.  This means that there are ``ghost''
D3-branes with the opposite charge {\it and tension} of a normal D3-brane.
Formally the force between ghost D3-branes and normal D3-branes vanishes,
but the kinetic terms for fluctuations of a ghost D3-brane are negative.
This is apparently invisible in five-dimensional supergravity; nevertheless
I regard non-positive $\sigma$ as unphysical.  A second point to observe is
that the geometry away from the $n=6$ brane distribution, (\ref{D3Gen}), is
invariant if we change all the $\ell_i^2$ by the same additive constant.
In particular, when all the $\ell_i$'s are equal, the geometry outside the
$\delta'$-function shell in the $n=6$ distribution is perfect $AdS_5 \times
S^5$.  This continues to hold when we replace the $\delta'$-function with
an ordinary $\delta$-function (thus curing the pathology of ghost
D3-branes).  This geometry was considered in
\cite{ChepelevRoiban,GiddingsRoss}.  The only gauge singlets that have
VEV's are also $SO(6)$ singlets, the simplest being $\tr \sum_I X_I^2$.
These operators are not in short multiplets, and are expected to acquire
large anomalous dimensions in the strong coupling limit.  How this squares
with the exactness of the classical moduli space of ${\cal N}=4$
super-Yang-Mills is a source of some puzzlement for me.

Extracting the distributions (\ref{d3distfun}) from the explicit
five-dimensional geometries that N.~Warner discussed is a slightly involved
calculation.  I will summarize only the most interesting part: the
consistent truncation ansatz.

Consistent truncation is often mistaken for the imprecise statement that
all but finitely many Kaluza-Klein modes can be consistently set to zero in
reduction of supergravity theories on compact manifolds.  In the case of
$S^5$ it is difficult to see how this can be so, since both the modes that
are kept and the modes that are cast out include $SO(6)$ singlets and
non-singlets.  A more straightforward way to think about consistent
truncation is as a technique for generating exact solutions to a higher
dimensional supergravity from the data of a lower-dimensional one.

Unlike for eleven dimensional supergravity on $S^7$
\cite{deWitOne,deWitTwo} or $S^4$ \cite{stny}, the full solution-generating
ansatz for ten-dimensional type IIB supergravity on $S^5$ is not known even
in implicit form.  We do however have the full ansatz when only scalars are
excited; essentially it was stated in \cite{KPW}.  Briefly, the $42$
scalars in the ${\cal N}=8$ $d=5$ supergraviton multiplet can be
parametrized by a 27-bein ${\cal V}_{AB}{}^{ab}$ for the coset
$E_{6(6)}/USp(8)$.  The indices $a$ and $b$ are in the fundamental ${\bf
8}$ of $USp(8)$.  In ${\cal V}_{AB}{}^{ab}$ they are antisymmetrized and
the symplectic trace is removed: this results in the $27$-dimensional
representation of $USp(8)$, which is also the fundamental representation of
$E_{6(6)}$.  A common convention is to use $z^{AB}$ to denote a vector in
the ${\bf 27}$ of $E_{6(6)}$ and $z_{AB}$ for the $\overline{{\bf 27}}$.

A crucial decomposition in the construction of $d=5$ ${\cal N}=8$ gauged
supergravity \cite{GRW} is
  \eqn{ESixDecompose}{\eqalign{
   E_{6(6)} &\supset SL(6,{\bf R}) \times SL(2,{\bf R}) \cr
   27 &= (\overline{{\bf 6}},{\bf 2}) + ({\bf 15},{\bf 1})  \cr
   z^{AB} &= (z^{I\alpha},z_{[IJ]}) \ ,
  }}
 where we have indicated in the second and third lines the way the ${\bf
27}$ of $E_{6(6)}$ splits under this decomposition.  Lowered indices $I$,
$J$ refer to the ${\bf 6}$ of $SL(6,{\bf R})$.  They key fact in
\ESixDecompose\ which makes it possible to gauge the obvious $SO(6)$
subgroup of $SL(6,{\bf R})$ is that the ${\bf 27}$ of $E_{6(6)}$ contains
the adjoint of $SO(6)$ (it is the ${\bf 15}$).  That is important because
the vector fields of ungauged $d=5$ ${\cal N}=8$ supergravity transform in
the ${\bf 27}$.  These vectors correspond (roughly) to components of the
metric with one leg in the non-compact dimensions and one along a Killing
vector, $K^{m\,IJ}$, of $S^5$.  (Here I use $m$ to denote a tangent space
index to $S^5$; $IJ$ label the relevant element of the ${\bf 15}$).  Up to
a normalization, the $K^{m\,IJ}$ can be taken to be the projections onto
$S^5$ of the differential operators $x_I \partial_J - x_J \partial_I$.

Now suppose we have some arbitrary solution to $d=5$ ${\cal N}=8$ gauged
supergravity involving only the metric, $ds_5^2$, and the scalars, ${\cal
V}_{AB}{}^{ab}$.  To construct the ten-dimensional Einstein metric, we
proceed as follows.  Define $K^m{}_{ab} = K^{m\,IJ} ({\cal
V}^{-1})_{IJ\,ab}$.  Solve the simultaneous equations
  \eqn{ScalarAnsatz}{\eqalign{
   \Delta^{-2/3} \tilde{g}^{mn} &= K^m{}_{ab} K^n{}_{cd} \Omega^{ac}
    \Omega^{bd}  \cr
   \Delta^2 &= \det(\tilde{g}_{mn} g_{(0)}^{np})
  }}
 for $\Delta$ and the metric $\tilde{g}_{mn}$ (whose inverse is
$\tilde{g}^{mn}$).  The metric $g_{(0)\,mn}$ is the usual round metric on
$S^5$.  Then the ten-dimensional metric is
  \eqn{TenMetric}{
   ds_{10}^2 = \Delta^{-2/3} ds_5^2 + \tilde{g}_{mn} d\psi^m d\psi^n \ ,
  }
 where $\psi^m$ are coordinates on $S^5$.  General elements of
$E_{6(6)}/USp(8)$ can lead to metrics $\tilde{g}_{mn}$ which are difficult
to describe concisely.  However for the subspace $SL(6,{\bf R})/SO(6)$
parametrized by the 20 scalars dual to $\tr X_{(I} X_{J)}$, $\Delta^{-2/3}
\tilde{g}_{mn}$ is the metric of an ellipsoid.

Once the metric \TenMetric\ has been obtained, it is relatively
straightforward to find appropriate coordinates $y_I$ in which the metric
takes the form \MultiCenter.

\section{Two-point functions and the spectrum of supergravity excitations}
\label{TwoPoint}

Given a two-point function $\langle {\cal O}(x) {\cal O}(0) \rangle$, there
is a straightforward way of extracting the spectrum of excitations in the
field theory that can be excited using the operator ${\cal O}$.  Let us
define
  \eqn{PiDef}{
   \Pi(s) = \int d^4 x \, e^{ip\cdot x} 
     \langle {\cal O}(x) {\cal O}(0) \rangle \ ,
  }
 where $s = p^2$.  The function $\Pi(s)$ will be analytic in the complex
$s$-plane except possibly for branch cuts and/or poles on the real axis.
The discontinuity across the real axis, $\Disc \Pi(s) = \Pi(s+i\epsilon) -
\Pi(s-i\epsilon)$, is the spectral measure of the invariant masses of the
states which ${\cal O}(0)$ can create from the vacuum.  If we use the
AdS/CFT prescription for computing Green's functions
\cite{gkPol,witHolOne}, which reads, schematically,
  \eqn{GreenPrescribe}{
   W_{CFT}[\phi_0] = \log \left\langle \exp \int d^4 x \, \phi_0 {\cal O} 
    \right\rangle_{CFT} = 
      \mathop{\rm extremum}_{\displaystyle{\phi 
       \mathop{\longrightarrow}_{\partial AdS} \phi_0}} S_{SUGRA}[\phi] \ ,
  }
 what one finds is that this spectral measure is precisely the spectrum of
the linearized equation for excitations of $\phi$ in the bulk spacetime.

\begin{figure}
   $$\vcenter{\openup1\jot
     \halign{# & # \cr 
    $V$ & $V$  \cr\noalign{\vskip1\jot}
    \ \psfig{figure=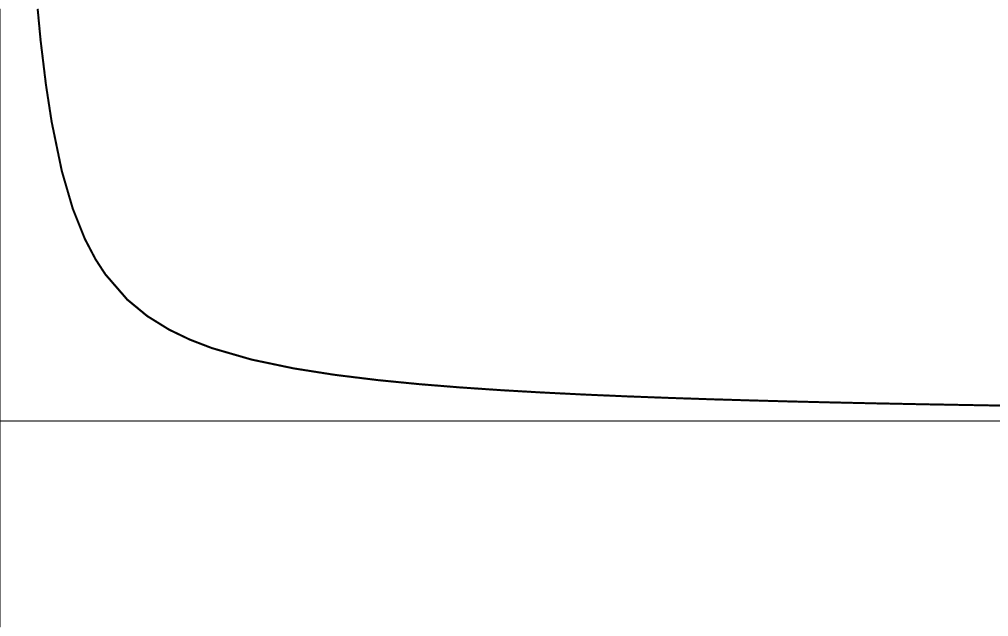,width=1.5in} \ 
     \ooalign{\raise20pt\hbox{$z$}} & 
    \ \psfig{figure=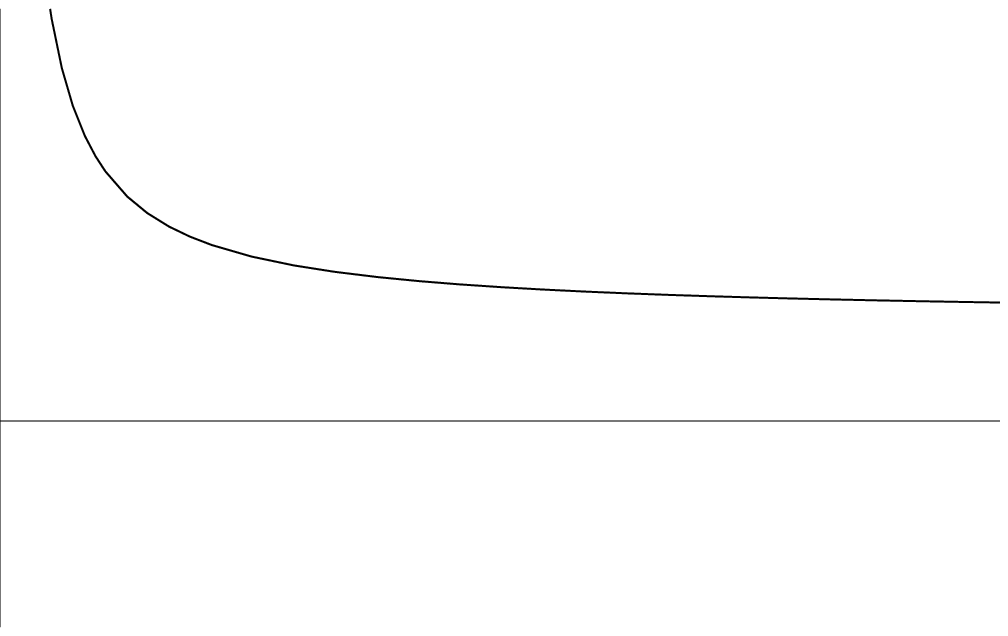,width=1.5in} \
     \ooalign{\raise20pt\hbox{$z$}}  \cr\noalign{\vskip-2\jot}
     \hfil {\Large a)} \hfil & \hfil {\Large b)} \hfil \cr\noalign{\vskip4\jot}
    $V$ & $V$  \cr\noalign{\vskip1\jot}
    \ \psfig{figure=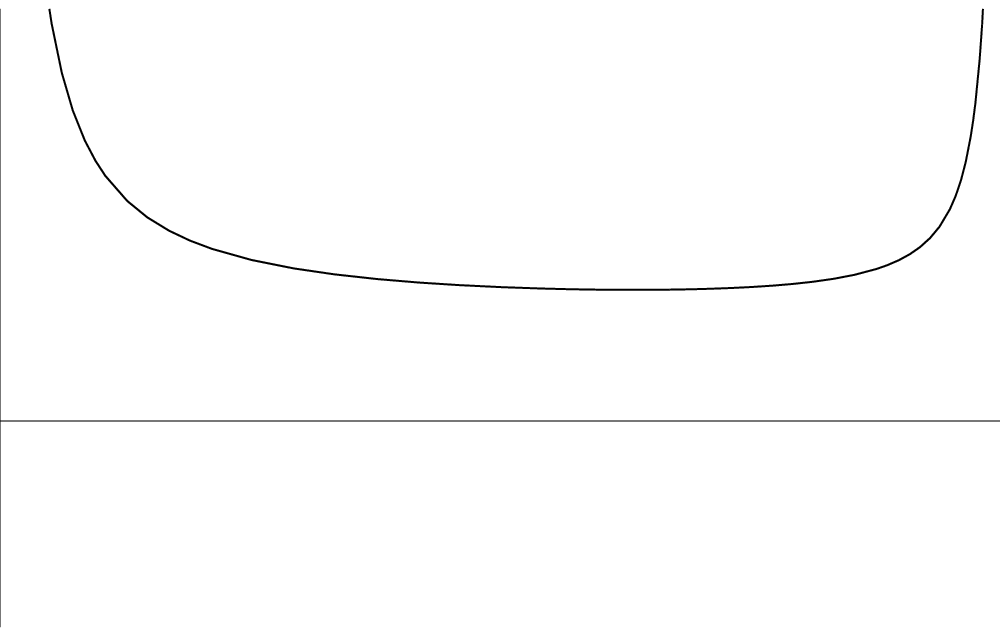,width=1.5in} \
     \ooalign{\raise20pt\hbox{$z$}} &
    \ \psfig{figure=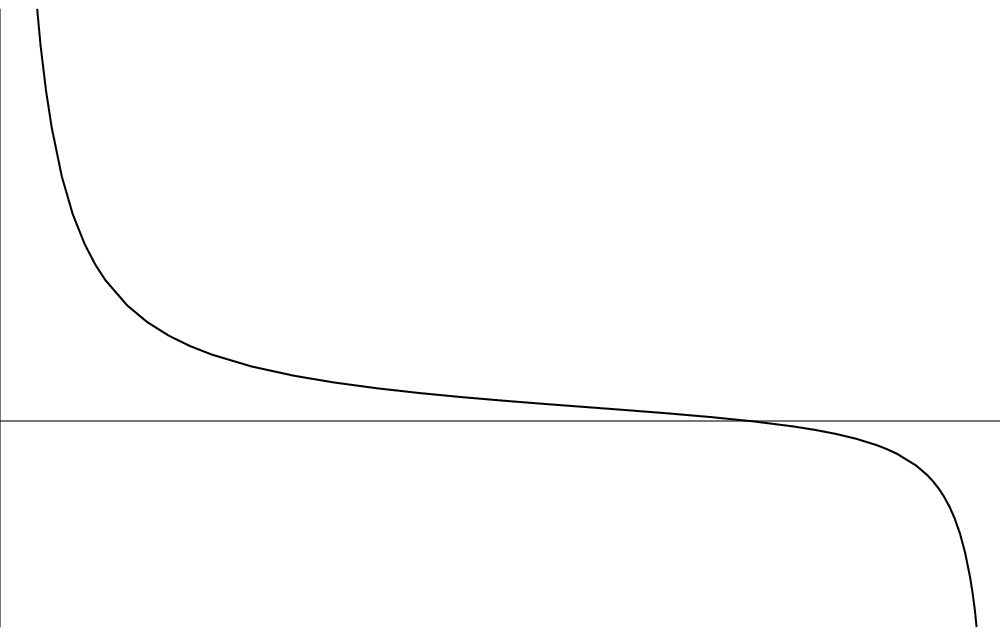,width=1.5in} \
     \ooalign{\raise20pt\hbox{$z$}}  \cr\noalign{\vskip-2\jot}
     \hfil {\Large c)} \hfil & \hfil {\Large d)} \hfil  \cr
  }}$$
 \caption{The various behaviors for $V(z)$ far from the boundary of
$AdS_5$: a) Vanishes; b) Asymptotes to a finite value; c) Increases without
bound; d) Decreases without bound.}\label{figB}
  \end{figure}

To be more explicit, consider the example of the five-dimensional dilaton,
which is dual to the dimension four operator ${\cal O}_4 = \tr (F^2 +
\bar\psi \slashed{\partial} \psi - X \square X + \ldots)$.  (I have not
been fastidious about numerical factors here).  The five-dimensional wave
equation is $\square\phi = 0$.\footnote{Actually all the supergravity
calculations can be done equally efficiently starting from the
ten-dimensional backgrounds.  The five-dimensional dilaton is the $s$-wave
of the ten-dimensional dilaton.}  This can be solved using separation of
variables: defining a radial variable $z$ such that the five-dimensional
metric takes the conformally flat form
  \eqn{ConformalMetric}{
   ds^2 = e^{2 A(z)} (dt^2 - dx_1^2 - dx_2^2 - dx_3^2 - dz^2) \ ,
  }
 we find 
  \eqn{Separate}{\eqalign{
   &\phi = e^{-i p \cdot x} e^{-3 A(z)/2} R(z)  \cr
   &[-\partial_z^2 + V(z)] R(z) = p^2 R(z)  \cr
   &V(z) = {3 \over 2} A''(z) + {9 \over 4} A'(z)^2 \ .
  }}
 The spectrum the Schroedinger operator $[-\partial_z^2 + V(z)]$ is
precisely the spectrum of excitations created by ${\cal O}_4$.  The
expression for $V$ in the last line of \Separate\ has the form of
supersymmetric quantum mechanics, so we know that the spectrum is bounded
below, $p^2 \geq 0$.\footnote{My lecture at Strings '99 and the initial
versions of \cite{fgpwTwo} were slightly mistaken on this point.  The error
was caught in \cite{cglp}.}  The qualitative behavior of the spectrum can
be reasoned out directly from the shape of $V(z)$.  The possibilities are
shown in figure~\ref{figB}.
 The first is encountered for the $n=1$ flow, and the spectrum for $p^2$ is
$(0,\infty)$.  The second is encountered for $n=2$, and the spectrum is
$(\ell^2/L^4,\infty)$.  The third is encountered for $n=3$, and the fourth
for $n=4,5$; in these cases the spectrum is discrete.  It is in fact
calculable for $n=4$, with the result $p^2 = {4 \ell^2 \over L^4} j(j+1)$
for $j=1,2,3,\ldots$.

A more precise and exhaustive description of the spectrum and the
explicitly calculable two-point functions can be found in
\cite{fgpwTwo,ABKS,cglp}.  The five- and ten-dimensional geometries were
also studied in \cite{BakasSfetsos}, and the linearized fluctuation
equations were related to integrable models of the Calogero-Moser type.

The outstanding puzzle that emerges from the analysis of the spectrum is,
how is Coulomb branch physics giving rise to mass gaps and discrete
spectra?  After all, the Coulomb branch is distinguished in that there are
long range forces, which suggests massless excitations.  I do not have a
good answer, but I will make some remarks which tend to make the obvious
answers less plausible and conclude with a (perhaps wild-eyed) suggestion
of what might be going on.

The first obvious point is that there are only $O(N)$ truly massless modes,
corresponding to the un-Higgsed $U(1)^{N-1}$ gauge group.  We do not expect
classical supergravity to see subleading effects in $N$.  Thus perhaps the
massive excitations we are seeing are color singlet combinations of the
$O(N^2)$ massive modes, corresponding to strings stretched between
different D3-branes in the distribution.  In some sense I think this idea
has to be right.  But there are some difficulties left.  First, the typical
mass of a stretched string is $\ell/\alpha'$, where $\ell$ is the diameter
of the brane distribution.  But in supergravity the scale of the mass gap
(or the discrete spectra) is instead $\ell/L^2 = \ell/(\alpha'
\sqrt{g_{YM}^2 N})$.  The stretched strings are BPS states, so their masses
are protected; but color-singlet bound states of them need not be BPS, and
their masses could be smaller than the sum of the masses of the
constituents because of binding energy.  In fact, maybe the binding energy
lowers the mass by the requisite factor $\sqrt{g_{YM}^2 N}\,$.\footnote{To
my knowledge this was first proposed by J.~Polchinski.}  In a lot of ways
this sounds like the most plausible way out of our difficulties.  However I
am not wholly satisfied with it for the following reasons.  First, the BPS
mass spectrum of stretched strings extends continuously to $0$ in the large
$N$ approximation.  A fraction on the order $(g_{YM}^2 N)^{-(1+\delta)/2}$
of the stretched strings have mass less than the supergravity mass gap,
$\ell/L^2$.  For all the positive definite mass distributions ($n \leq 4$
in (\ref{d3distfun})) we have $\delta \leq 2$, so this is still a large
number of states.  Binding energy by definition cannot be positive, so the
existence of these states seems to contradict the existence of a mass gap.
(One way out of this is to say that because their number is suppressed by a
power of the 't Hooft coupling, supergravity simply misses them.  It would
be interesting to estimate more carefully the $\alpha'$ corrections in the
regions of strong curvature and test whether this escape is plausible.)
Second, the $n=1$ distribution exhibited no mass gap in supergravity.  What
aspect of the field theory binding mechanism could be responsible for
producing a gap only when $n>1$?

A possible resolution to the second question is that what supergravity
measures is not simply the correlator $\langle {\cal O} {\cal O} \rangle$
in a particular state on the Coulomb branch, but rather the averaged
quantity $\overline{\langle {\cal O} {\cal O} \rangle}$, where the
averaging is over all discrete distributions of $N$ D3-branes which are
consistent with the continuous distribution $\sigma$.  Say $\sigma$ is a
$p$-dimensional distribution in the transverse $6$ dimensions: $p=1$ for
the case $n=1$, where the D3-branes are distributed along a line segment;
$p=2$ for $n=2$, where the D3-branes are distributed over a disk; and $p=3$
for $n=3$ and $n=4$, where the branes are distributed in a 3-dimensional
ball or across the surface of an $S^3$.  Intuitively speaking, the branes
are allowed to move a distance comparable to the nearest-neighbor distance,
$\ell/N^{1/p}$.  The suggestion advanced in \cite{fgpwTwo} is that
averaging over the ensemble of brane distributions that we get by allowing
individual D3-branes to ``wiggle'' a distance $\ell/N^{1/p}$ leads to an
altered lagrangian which includes trace-squared terms.  The guess is that
these terms are sufficiently dominant for $n>2$ to give rise to a discrete
spectrum, but are insignificant for $n<2$.  A naive estimate is that their
coefficient would go as $N^{1-2/p}$, as compared to $1$ for the usual terms
in the ${\cal N}=4$ lagrangian.  I would emphasize however the speculative
nature of this idea.  No successful calculation has been carried out to
verify that ensemble averaging could change the behavior of the Green's
functions so radically.

\section{Conclusions}

The one universal result in non-conformal examples of the AdS/CFT
correspondence is the c-theorem.  In supergravity it says that the function
$C(r) = {1 \over A'(r)^{d-1}}$ is monotonic in $r$ for the metric
\FiveGeometry\ in $d+1$ non-compact bulk dimensions.  This is simply a
consequence of the null energy condition, which is satisfied for all types
of matter that arise in string theory (with the possible exception of
orientifold planes, where there is a formally negative tension).  In field
theory, $C(r)$ serves as a useful approximate guide to the number of
degrees of freedom.  It is approximate partly because supergravity only
knows about the leading order in $N$, but mostly because the correspondence
between radius and energy scale is not known precisely except in the case
of pure $AdS_5$.  When the geometry approaches $AdS_5$, $C(r)$ is a
constant proportional to the trace anomaly coefficients in the conformal
field theory.  This and its monotonicity make it the best candidate we have
for a c-function in the AdS/CFT context.

There is a five-parameter family of states on the Coulomb branch of ${\cal
N}=4$ super-Yang-Mills theory which admits a dual description wholly in
terms of $d=5$ ${\cal N}=8$ gauged supergravity.  These are states where
the VEV's of the operators $\tr X_{(I} X_{J)}$ are adjusted arbitrarily.
The VEV's of all the other symmetric traceless products can best be
calculated by raising the five-dimensional solution up to ten dimensions
via the consistent truncation ansatz.  The five-dimensional geometries
provide examples of ``flows'' where $C(r)$ goes to zero, either smoothly as
$r \to -\infty$ or sharply at a finite value of $r$.  We understand this
qualitatively as reflecting the fact that at lower energies, fewer of the
massive stretched string excitations are available.  However the
characteristic energy scale of correlators calculated in the supergravity
geometry is $\ell/L^2$: smaller by a factor of $\sqrt{g_{YM}^2 N}$ than the
characteristic energy scale of the Higgs VEV's which define the Coulomb
state.  Strong gauge interactions could be responsible for color singlet
combinations of BPS stretched string states giving up most of their mass to
binding energy.  I have suggested that averaging over ensembles of brane
distributions may also play a role in determining the qualitative features
of the spectra and their remarkable dependence on the dimension of the
distribution.

\section*{Acknowledgements}

I thank D.~Freedman, N.~Warner, and K.~Pilch for a very stimulating
collaboration.  I also thank S.~Giddings, R.~Myers, V.~Periwal,
J.~Polchinski, and M.~Porrati for useful discussions.  The organizers of
Strings '99 have my gratitude for putting together a remarkable conference.
The research reported on in this lecture was supported in part by the
Harvard Society of Fellows, by the NSF under grant number PHY-98-02709, and
by DOE grant DE-FGO2-91ER40654.

\bibliography{talk}
\bibliographystyle{ssg}

\end{document}